\documentclass[11pt]{article}
\usepackage{amsmath,amssymb,amsfonts,latexsym,graphicx}
\numberwithin{equation}{section}
\newcommand{\be}{\begin{equation}} \newcommand{\ee}{\end{equation}}
\newcommand{\beq}{\begin{equation}} \newcommand{\eeq}{\end{equation}}
\newcommand{\bea}{\begin{eqnarray}} \newcommand{\eea}{\end{eqnarray}}
\newcommand{\field}[1]{\mathbb{#1}}
\title{
\begin{center}
 { \bf IIA moduli stabilization with badly broken supersymmetry.}
\end{center}
}

\begin{document}

\maketitle
\renewcommand{\thefootnote}{\fnsymbol{footnote}}
\centerline{{\bf  Michael Dine, Alexander Morisse, Assaf Shomer
and Zheng Sun \footnote{\tt dine@scipp.ucsc.edu,
amorisse@physics.ucsc.edu, shomer@scipp.ucsc.edu;
zsun@physics.ucsc.edu}}}

\centerline{\it
Santa Cruz Institute for Particle Physics}
\centerline {\it 1156 High Street, Santa Cruz, 95064 CA, USA }

\begin{abstract}
Scherk-Schwarz compactification in
string theory can be defined as orbifolding by an R symmetry,
a symmetry that acts
differently on bosons and fermions. Such a
symmetry can arise in many situations, including toroidal and
orbifold compactifications, as well as smooth Calabi-Yau spaces.
If the symmetry acts freely then for large radius there are no
tachyons in the spectrum. We focus mainly on stabilization
by fluxes, and give examples with all moduli stabilized where the coupling is
small and the internal manifold is large.  Such models appear to be perturbatively stable with
supersymmetry broken at the Kaluza-Klein scale. These are
interesting laboratories for a variety of theoretical questions
and provide models of a non-supersymmetric landscape.
\end{abstract}
\newpage

\setcounter{footnote}{0}
\renewcommand{\thefootnote}{\arabic{footnote}}

\section{Introduction}

One of the simplest ways to think about supersymmetry breaking in
higher dimensions is Scherk-Schwarz compactification \cite{SS}.
Here one usually considers compactification on a torus, and
imposes periodic boundary conditions for bosons and anti-periodic
boundary conditions for fermions.  As a result, in the four
dimensional theory, supersymmetry is broken at the scale of
compactification, $1/R$; below this scale, there is no light
gravitino, and the breaking
appears explicit. These theories have other interesting features.
Typically, at small radius, they have tachyons and are T-dual to
compactifications of Type $0$ theories \cite{sw}. In addition,
even in the tachyon-free regime, these compactifications suffer
from the Witten Kaluza-Klein instability \cite{wittenkk}. The
significance of these features is hard to assess, however, due to
the perturbative instabilities which arise already at one loop.
Generally, it is not clear whether any of these classical
solutions correspond to stable or metastable states of the quantum
theory.

The term ``Scherk-Schwarz compactification" is most often used for
these toroidal compactifications, but, the concept is more general\cite{kounnas,kiritsis,
angelatonj}.   As we will review and elaborate in this
paper, it should be applied to a broader array of models.  This
more general set of constructions is obtained by modding out a
string solution by a freely acting symmetry operation that acts
differently on bosons and fermions, namely, a freely acting R
symmetry\footnote{As we explain below, in practice these $R$
symmetries are $Z_2$'s times ordinary
symmetries.}$^,$\footnote{One can also generalize the
Scherk-Schwarz projection in other directions. For example one can
consider a combination of a U-duality transformation and a
translation \cite{hull}, \cite{schulz}.}. These symmetries can
include rotations in compact or in non-compact directions as well
as abstract symmetries of conformal field theories. Modding out by
these symmetries can be thought of as including Wilson lines for
the spin-connection, which break supersymmetry. From the point
of view of the resulting string models, there is no fundamental difference
between freely interacting symmetries and those with fixed points.
The reason we
confine our attention to symmetries that are freely acting is to
ensure that at large radius there are no tachyons or new light
states in twisted sectors. Even then, there are often tachyons at
small radius, as one would predict from the dualities alluded to
above.  Even though the effective theory below the compactification
scale has no symmetry, the moduli space, classically, consists of the entire
subspace invariant under the symmetry.

In Scherk-Schwarz Models, the vacuum energy receives corrections
already at one loop\cite{rohm}. In a geometrical compactification,
the effective potential will typically drive the radius towards
smaller values, where tachyons start to appear in the spectrum.
Recently, however, there has been much progress in constructing
stable or metastable string ground states in supersymmetric (and
nearly supersymmetric) models \cite{bp}-\cite{douglasdenef}.
Crucial to these constructions is the role of fluxes in
stabilizing moduli. Much attention has been focussed on
supersymmetric compactifications of (orientifolded) IIB theories,
where fluxes fix complex structure moduli already at the level of
a classical analysis. Kahler moduli then are often fixed by
non-perturbative effects\cite{kklt}. The resulting spaces can be
dS or AdS.

As we will explain, one can repeat the IIB construction with a
Scherk-Schwarz projection for many Calabi-Yau manifolds.  It is
only necessary that the original Calabi-Yau possess a suitable R
symmetry on some subspace of its moduli space, and to set to zero all
fluxes which transform non-trivially under the symmetry.  At the
classical level, the potential for the moduli which survive the
projection is the same as in the model before the projection; the
absence of fixed points insures the absence of new massless
particles.  So again, one can find examples where all of the
complex structure moduli can be fixed by fluxes.  However, the
Kahler moduli will receive corrections already in perturbation
theory, and any minimum of their potential is likely to reside at
small radii (radii of order the string scale), so one can at best
speculate about the possible existence of metastable states (much
as in \cite{douglasdenefnonsusy}).

An alternative possibility for fixing moduli is  provided by a
recent analysis of IIA compactifications with
fluxes\cite{dewolfe}.  In this case, it has been argued that, for
suitable fluxes, one can stabilize all of the moduli classically.
More dramatically, it appears that there are an infinite series of
such states, with arbitrarily small string coupling and curvature.
Moreover, the geometry of the solutions turns out to be a product
of 4 dimensional AdS space times a compact CY manifold with an
arbitrarily large hierarchy between the AdS curvature radius and
the KK scale, so that unlike Freund-Rubin
compactifications\cite{freundrubin}, the physics is well captured
by a 4 dimensional effective description. A subset of these states
are supersymmetric.  The rest are {\it approximately
supersymmetric}, in the sense that the {\it gravitino mass is
parameterically small compared to the Kaluza-Klein (KK) scale}. In
this paper we will show that one can consistently apply a
generalized Scherk-Schwarz projection to these models, thus
constructing, with the same level of reliability, an infinite
sequence of states with {\it badly broken supersymmetry}, namely,
where supersymmetry is broken at the KK scale. From a four
dimensional viewpoint, such generalized Scherk-Schwarz models are
non-supersymmetric. There is no scale at which the theory appears
four dimensional and is (even approximately) supersymmetric. These
states provide an interesting laboratory in which to study a
variety of questions, including perturbative and non-perturbative
stability and statistics of non-supersymmetric states.

This paper is organized as follows.  In section \ref{ssrev}\ we
present a brief review of the standard Scherk-Schwarz projection
\cite{SS}\ its application in string theory \cite{rohm}\ and
briefly explain the duality to type $0$ strings. We describe in
particular how the usual Scherk-Schwarz projection can be phrased
as modding out by a freely acting discrete R-symmetry.  This is
particularly transparent in the Green-Schwarz formulation; a more
detailed analysis in the RNS formalism appears in an appendix. In
section \ref{genss}\ we consider a variety of generalizations of
Scherk-Schwarz. In section \ref{2ass}\ we present a generalized
Scherk-Schwarz projection on the type IIA $T^6/\field{Z}_3^2$ CY
orientifold of DeWolfe et.al \cite{dewolfe}. We show how the
classical stability analysis carried in \cite{dewolfe}\ remains
intact even with broken supersymmetry, and give a formal argument
that quantum correction are parameterically small.
\footnote{Recent work calls into question the possibility of a
systematic weak coupling analysis\cite{banks}.} In the concluding
section, we remark on some possible applications of these ideas.
These include developing a more general understanding of Witten's
bubble of nothing\cite{wittenkk}, and the use of
non-supersymmetric models with stabilized moduli to study
questions about the landscape. Two appendices present an RNS
formulation of the standard Scherk-Schwarz projection and an
analysis of a $T^6/\field{Z}_4^3$ CY orientifold.

There are other approaches to constructing models of broken
supersymmetry in string theory which lead to models with some
similar features, such as supersymmetry broken at the Kaluza-Klein
scale (see, for example, \cite{silverstein}, where
compactification on products of Riemann surfaces leads to models
with badly broken supersymmetry). The reason for our focus on
Scherk-Schwarz models lies in their simple realizations,
classically, as critical string theories, and in some cases
quantum mechanically as small distortions of such theories.

\section{Scherk-Schwarz models in field theory and string theory}\label{ssrev}

Consider a string theory
compactified on one periodic dimension:
\begin{equation}\label{arr}
X\sim X+2\pi R
\end{equation}
with left and right moving momenta given by
\begin{equation}\label{plpr}
p_L=\frac{m}{R}+\frac{wR}{2},\quad p_R=\frac{m}{R}-\frac{wR}{2}
\end{equation} where $m,w$ are integral momentum and winding quantum numbers, and we set $\alpha^{\prime}=2$.
E.g, the energy of a superstring excitation on $R^{1,8}\times
S_X^1$ in the sector with winding number $w$ and with $m$ units of
momentum and with left (right) oscillator numbers $N\ (\tilde N)$
is
\begin{equation}\label{mass}
\begin{split}
m^2=2(N-\frac{1}{2})+p_L^2=2(\tilde N-\frac{1}{2})+p_R^2=(N+\tilde N-1)+\bigl(\frac{m}{R}\bigr)^2+\bigl(\frac{wR}{2}\bigr)^2\\
\end{split}\end{equation}
subject to the level matching condition
\begin{equation}\label{lm}
 N-\tilde N+mw=0.
 \end{equation}

In a Scherk-Schwartz compactification \cite{SS}\ one imposes anti-periodic boundary conditions for the spacetime fermions along the periodic coordinate $X$. Consequently, fermions have half integral momenta along the circle parameterized by $X$. Supersymmetry is spontaneously broken as the bosons and fermions have different mode expansions along the circle which translate to different energies in the dimensionally reduced model. This fact is true already in field theory.

In string theory there are additional features \cite{rohm}. Imposing anti-periodic boundary conditions for fermions around the circle is equivalent to modding out by the $\field{Z}_2$ R-symmetry $e^{2\pi i R P^X} (-1)^F$. In order to obtain a modular invariant partition function one needs to add the twisted sectors. These are particularly easy to understand in the Green-Schwarz formulation\footnote{A useful trick is to think of this as a geometrical orbifold on a circle with a doubled radius $2R$. The twisted states wind half-way around that circle and acquire anti-periodic boundary conditions.} where the space-time fermion number operator reverses the sign of the two dimensional fields $S_a$.
\begin{equation}\begin{split}
X(\sigma + \pi) &= X(\sigma) + w\cdot 2\pi R\\
S_a(\sigma + \pi) &= (-1)^w S_a(\sigma).\\
\end{split}\end{equation}
The ground state energy in odd winding sectors is thus
\begin{equation}
m^2=\bigl(\frac{wR}{2}\bigr)^2-1
\end{equation} which becomes tachyonic for small enough radius. Also, this projection effectively reverses the familiar type $II$ GSO projection between odd and even winding sectors\footnote{A simple derivation of this fact appears in appendix A}.
Thus, by imposing Scherk-Schwarz boundary conditions in type $II A/B$, the bosonic spectrum is given by (using the notations of Polchinski \cite{pol})
\begin{itemize}
\item $w=2w^{\prime}\ $: $(NS+,NS+), (R+,R\pm)$
\item $w=2w^{\prime}+1\ $: $(NS-,NS-), (R-,R\mp)$.
\end{itemize}

\subsection{Duality with type $0$}
Scherk-Schwarz compactifications of type II superstrings obey
a certain T-duality relation with type 0 string theory.
In the limit that $R\rightarrow 0$ all the spacetime fermions
become very massive because they can not have
zero modes along the circle. It can be shown\footnote{More details appear
in appendix B.} that type IIA/B compactified on
a circle of radius $R$ with Scherk-Schwarz boundary
conditions, in the limit $R\rightarrow 0$, is equivalent
to type $0 B/A$ in uncompactified spacetime.

Calabi-Yau spaces are classical solutions of the Type 0 theories.
One can see this in two ways, which will be useful for us in what
follows.  First, at the classical level, the field equations for
the bosonic fields are the same as in the Type II theories.
Therefore, classically, solutions of the two theories coincide.
Alternatively, a compactification of Type II on a
Calabi-Yau space is described by two dimensional
left and right $\mathcal{N}=2$ SCFT with $c=9$. This
CFT can always be used also as a sensible background of Type 0.
Thus, there are large classes of smooth manifolds which solve the
classical equations of the Type 0 theory. One can even introduce
fluxes, and classically, the potential for these will be as in the
Type II theories. But these constructions do not seem terribly
interesting, since at large radius they possess tachyons (at least
at weak coupling), and quantum corrections will generally
destabilize them. In the next section, we will consider
generalized Scherk-Schwarz constructions with the potential to
avoid these difficulties.

\section{Generalizing Scherk-Schwarz}\label{genss}

Modding out a string theory by an $R$ symmetry will yield a string
configuration with less supersymmetry. Supersymmetry will be
broken at the KK scale associated with the compact manifold.

In this section we provide several examples of this kind of
generalized Scherk-Schwarz projections. We start with toroidal and
toroidal orbifold models. These are familiar models described by
free two dimensional fields. Then we note that large classes of
Calabi-Yau compactifications, both of Type II and heterotic
strings, admit such projections. These are distinctly less
trivial. In the next section, we will show that fluxes can stabilize some or all of the
moduli of such compactifications.

\subsection{Scherk-Schwarz as an orbifold.}

The usual Scherk-Schwarz projection is a special case of the general
toroidal orbifold construction. As explained in
\cite{orbiII}\ a class of orbifolds can be constructed as a
quotient of the Euclidean space $\field{R}^d$ by a subgroup of
rotations and translations $g=(\theta, v)$ called the {\it
space-group} $S$. The orbifold is the quotient $\Omega=\field{R}^d/S$ where the elements of $S$ act on a vector $x\in\field{R}^d$ as
$gx=\theta x+v$. The subgroup $\Lambda$ of translations $(1,v)\in S$ is called {\it
the lattice of S}. The subgroup of $O(d)$ of rotations $\theta$ such that $(\theta,
v)\in S$ is called the {\it point-group}. It can be shown that
each element of the point group is associated with a unique vector
$v$ (up to lattice translation). The point-group is also the
holonomy group of the orbifold. An orbifold with a space-group
that has a non-trivial point-group can break supersymmetry because
bosons and fermions transform differently under rotations.

Scherk-Schwarz compactifications have a natural description in
this language as modding out by the space group generated by two
elements that include a  $2 \pi$ rotation and a translation in an orthogonal direction $\vec e$, namely,
$g=\{(e^{2\pi i\mathcal{J}_{ab}},0)\ ; (1,2\pi R\cdot
\vec{e})\}$.
More generally\cite{kounnas,kiritsis,angeltonj}, take any singular limit of a CY n-fold described by an
orbifold of n products $T^2\times\cdots\times T^2$.
Since the tori are smooth,
the holonomy matrix is just the identification
matrix of the orbifold $z_i\sim A_i^{\ \bar j}z_j$.
Demanding that $\det A=1$ gives a CY. However, if we demand instead that $\det A=e^{2\pi i}$ then the geometry does not change, but fermions will pick up a minus sign under the action of the orbifold.  This will break supersymmetry.
In general there will be tachyons in twisted sectors located at fixed
points of the orbifold. However, if we add a translation to the space
group so that the action is free, we can avoid tachyonic instabilities
at large radius. However, the radius modulus will generally tend to decrease in size until it enters the tachyonic regime. This can be avoided in a flux compactification. We give a detailed example of this procedure in section \ref{2ass}.

\subsection{Smooth Calabi-Yau Spaces in Type II and Heterotic Theories}

We can perform a Scherk-Schwarz projection in any Calabi-Yau space
which admits a freely acting symmetry.\footnote{In an earlier
version of this paper, we spoke of freely acting R symmetries.
However, Volker Braun has pointed out to us that in the case of
Calabi-Yau three-folds, freely acting symmetries always leave the
holomorphic three-form invariant (this follows from results stated
in \cite{gsw}).  As a result, any freely-acting $R$ symmetry must
be at most a $Z_2$, times some ordinary symmetry.}  This is
already possible for the familiar quintic in $CP^5$. Take, for
example, the quintic polynomial to be
\begin{equation}
P= z_1^5 + z_2^5 + z_3^5 + z_4^5 + z_5^5=0.
\end{equation}
Then the transformation
\begin{equation}
z_i \rightarrow \alpha^i z_i \qquad {\rm with}\qquad\alpha=e^{2\pi
i\over 5} \label{cp4transform}
\end{equation}
is a freely acting symmetry of the theory\cite{gsw}. As for the
orbifold case, we can define the action of this symmetry to flip
the sign of the supercharges (covariantly constant spinor), so
that supersymmetry is broken.  In this case, 21 complex structure moduli
and the one Kahler modulus survive the projection.

If one examines, say, lists of Calabi-Yau manifolds defined in weighted projective
spaces, one can find many suitable symmetries.  In all of these cases,
there remains a large moduli space.  Again, it should
be stressed that these are solutions of the classical equations, but quantum
corrections, already in string perturbation theory, will generate a potential
for the moduli.

It should be noted that these manipulations are valid in Type II and heterotic theories.
Heterotic theories actually yield a richer set of models, since in addition to
$(2,2)$ theories, one has $(2,0)$ theories.  Many of the latter admit suitable $R$ symmetries
as well\cite{miracles}.

\section{Scherk-Schwarz version of IIA models.}\label{2ass}

In this section we present a Scherk-Schwarz version of the
type IIA superstring model analyzed by DeWolfe et.al in
\cite{dewolfe}, resulting in a sequence of non-supersymmetric,
tachyon free models. We first review the supersymmetric construction,
and then modify it so as to implement a Scherk-Schwarz projection.
In an appendix, we study a different orbifold.

\subsection{Review of supersymmetric IIA Constructions}\label{2arev}

DeWolfe et al construct an infinite set of AdS states where the
string coupling and curvature can be made arbitrarily small by
choice of fluxes. They start with compactification of Type II
theory on the $\field{Z}_3$ orbifold \cite{Z3}.
This orbifold is a singular limit of a CY with
$\chi=72$ obtained by starting with a $T^6$ defined by the
identifications
\begin{equation}\label{trans}
z_i\sim z_i+1\sim z_i+\alpha,\quad i=1,2,3;\quad \alpha = e^{\pi i
\over 3}
\end{equation} and then modding out by a $\field{Z}_3$ symmetry of this lattice
defined by
\begin{equation}\label{TT}
z_i\rightarrow T_i^{\ j} z_j \qquad T = \left (
\begin{matrix}\alpha^2 & 0 & 0 \cr 0  & \alpha^2 & 0 \cr 0 & 0 &
\alpha^{-4} \end{matrix}\right)
\end{equation}
The choice of the phases in \ref{TT}\ ensures that the orbifold
has $SU(3)$ holonomy. This leaves, in the case of Type II
theories, $N=2$ supersymmetry in 4 dimensions. As was noted in
\cite{Z3}\ one can mod out by a further freely acting
$\field{Z}_3$
\begin{equation}\label{QQ}
z_i\rightarrow Q_i^{\ j} z_j+a_i\qquad Q=\left (
\begin{matrix}\alpha^2 & 0 & 0 \cr 0  & \alpha^{-2} & 0 \cr 0 & 0
& 1 \end{matrix}\right )\qquad a=\frac{1+\alpha}{3}\left (
\begin{matrix} 1 \cr 1 \cr 1 \end{matrix}\right )
\end{equation}
The last step reduces the supersymmetry down to $N=1.$

The orientifold projection involves the
symmetry
\begin{equation}\label{ori}
{\cal O} = \Omega_P (-1)^{F_L} \sigma
\end{equation}
with $\Omega_P$ world-sheet parity.  $\sigma$ is the reflection
\begin{equation}\label{sig}
\sigma:  z_i \rightarrow -\bar z_i
\end{equation}

The effect of these transformations is to reduce the supersymmetry to $N=1$ and'
to eliminate many of the moduli of the original toroidal compactification.
In the untwisted sector, only the diagonal moduli,
\begin{equation}
g_{i \bar i} = v_i, ~i=1,2,3.
\end{equation}
survive; they are each part of a chiral
multiplet, whose scalar components have the form
\begin{equation}
t_i = b_i + i v_i.
\end{equation}
There are nine twisted moduli, $t_A = b_A + i v_A$.

Now one introduces a number of fluxes.  There is a zero form flux. There are two-form
fluxes, three and four-form fluxes.  A basis of two-form fluxes, odd under the reflection
$\sigma$, is provided by:
\begin{equation}
\omega_i = (\kappa \sqrt{3})^{1/3} i dz^i \wedge  d\bar z^i,~ i=1,2,3.
\end{equation}
There is a corresponding set of four-cycles
\begin{equation}
\tilde w^i = \left ({3 \over \kappa} \right )^{1/3}
(i dz^j \wedge  d\bar z^j) \wedge (i dz^k \wedge  d\bar z^k).
\end{equation}
The three-form invariant under $T$ and $Q$ is:
\begin{equation}\label{omeg}
\Omega = e^{1/4} i dz_1 dz_2 dz_3 = {1 \over \sqrt{2}} (\alpha_0 + i \beta_0).
\end{equation}

DeWolfe et al turn on the fluxes:
\begin{equation}
H_3 = -p \beta_0 ~~~F_r = e_i \tilde \omega^i; ~F_0 = m_0.
\end{equation}
A simple analysis
leads to a potential for the moduli:
\begin{equation}\label{pot}
V= {p^2 \over 4} {e^{2\phi} \over {\rm vol}^2} + {1 \over 2}
(\sum_i^3 e_i^2 v_i^2) {e^{4 \phi} \over {\rm vol}^3}+ {m_0^2 \over 2}
{e^{4\phi} \over {\rm vol}} + \sqrt{2} m_0 p{e^{3\phi} \over
{\rm vol}^{3/2}}.
\end{equation}
Here ${\rm vol} = \kappa v_1 v_2 v_3$.
The potential has a local minimum:
\begin{equation}\label{min}
v_i = {1 \over \vert e_i \vert} \sqrt{{5 \over 3}
\vert {e_1 e_2 e_3 \over \kappa m_0} \vert}
~~~e^{\phi} = {3 \over 4} \vert p \vert
\left ( {5 \over 12} {\kappa \over \vert m_0 e_1 e_2 e_3 \vert }
\right ) ^{1/4}.
\end{equation}
DeWolfe et al also show that the moduli associated with the fixed points
can be stabilized as well.  They work out explicitly the geometry of the smooth manifold,
and show that, introducing suitable four-form fluxes associated with the fixed
points, the corresponding moduli have minima where the manifold is smooth.

Ref. \cite{dewolfe}\ then gave a purely four dimensional description of all
of this, in terms of a theory of light chiral fields, whose lagrangian is
described by a Kahler potential and a superpotential.  In this analysis, they
included two-form fluxes, $m_i$.  These can be used to tune the
axions (if the two form fluxes vanish, so do the axion fields).  We won't reproduce
their full equations for the superpotential, but will focus on the piece
involving the $t_a$ fields, and only in the special case that the $m_a$'s all vanish:
\begin{equation}
W^K(t_a) = e_0 + e_a t^a -{m_0 \over 6} \kappa_{abc} t_a t_b t_c.
\end{equation}
The Kahler potential for these fields is:
\begin{equation}
K^K(t_a) = -\log({4 \over 3} \kappa_{abc} v_a v_b v_c).
\end{equation}
Here the quantity $\kappa_{abc}$ is the triple intersection,
\begin{equation}
\kappa_{abc} = \int \omega_a \wedge \omega_b \wedge \omega_c.
\end{equation}

The resulting equations for the $v_a$'s are:
\begin{equation}
3 m_0^2 \kappa_{abc} v_b v_c + 10 m_0 e_a = 0.
\end{equation}
In the case of the $\field{Z}_3$ orbifold, these equations are simple to analyze.
This is because $\kappa_{abc}$ breaks up into distinct pieces involving
untwisted and twisted moduli.  In \cite{dewolfe}, an argument is given
for this based on the geometry, but there is another way to understand this
selection rule..
The model has a ``quantum symmetry"\cite{vafa}, a $\field{Z}_3$, under which the fields
in the sector twisted once by $T$ transform with a phase $e^{2\pi i/3}$, while
there antiparticles in the doubly-twisted sector transform by $e^{-2\pi i/3}$.
$\kappa$ determines the prepotential of the $N=2$ theory (before the orbifold
projection).  This potential is holomorphic in the fields, and must be invariant
under the $\field{Z}_3$.  As a result, only terms with zero or three twisted fields
are non-vanishing, and the equations for the untwisted and twisted moduli
decouple.  DeWolfe et al denote these by the indices $i$ and $A$, respectively, so
they are able to take
\begin{equation}
\kappa_{123}= \kappa; ~\kappa_{AAA} = \beta.
\end{equation}
In this way, they are able to find supersymmetric solutions, corresponding to
the solutions above, for certain choices of the signs of the $e_i$'s, and $f_A$'s
(the four-form fluxes at the fixed points):
\begin{equation}
v_i = {1 \over \vert e_i \vert} \sqrt{-5 e_1 e_2 e_3 \over 3 m_0 \kappa}
~~~~v_A = \sqrt{-10 f_A \over 3 \beta m_0}.
\label{vsolutions}
\end{equation}
The potential of eqn. \ref{pot} is quadratic in the $e$'s and $f$'s,
so the location of the minima is independent of the signs
of the fluxes.  But only for some choices of signs are there
solutions, as in eqn. \ref{vsolutions}; the minima for other choices
of flux are not supersymmetric.
But they are approximately so.  The gravitino mass is of order:
\begin{equation}
m_{3/2} = e^{K/2} W \approx {\vert e \vert^{-3/4}}\vert e \vert^{-3/2}
\vert e \vert^{3/2}\sim \vert e \vert^{-3/4}.
\end{equation}
Here the first $\vert e \vert^{-3/2}$ factor arises from the Kahler potential for the dilaton.
This is to be compared to the Kaluza-Klein scale, which is $1/\sqrt{v} \sim
\vert e \vert^{-1/4}.$
So the gravitino is isolated from the Kaluza-Klein tower, i.e. there is a range
of energy scales with a single, light gravitino, and the theory has an approximate
supersymmetry at low energies.

\subsection{A $T^6/\field{Z}_3^2$ orbifold with a generalized Scherk-Schwarz projection.}

The model of \cite{dewolfe} admits an immediate generalization.  These authors
performed two orbifold projections, both to reduce the supersymmetry to
$N=1$, and to obtain a comparatively small number of moduli.
We will also perform two projections, the first one is the same as in \cite{dewolfe}\ and the second is a slight variation which implements the Scherk-Schwarz projection. This will allow us to eliminate all supersymmetry, while obtaining essentially
the same set of moduli.

We thus start with the standard $Z_3$ orbifold described in the previous section in equations
\ref{trans}-\ref{QQ}.
Our generalized Scherk-Schwarz projection involves taking, in the
second step, a slightly different matrix:
\begin{equation}\label{tQ}
z_i\rightarrow \tilde Q_i^{\ j} z_j+a_i\qquad \tilde Q=\left (
\begin{matrix}\alpha^2 & 0 & 0 \cr 0  & \alpha^{4} & 0 \cr 0 & 0
& 1 \end{matrix}\right )\qquad a=\frac{1+\alpha}{3}\left (
\begin{matrix} 1 \cr 1 \cr 1 \end{matrix}\right )
\end{equation}
Gravitinos with zero momentum will pick up a minus sign, due to the overall $2\pi$
rotation in the internal tori, under this transformation. (Equivalently, under
parallel transport around the
non-contractible loop which in the original $T^6/\field{Z}_3$
orbifold connected the two fixed points at $(0,0,0)$ and
$\bigl((1+\alpha)/3,(1+\alpha)/3,(1+\alpha)/3\bigr)$, the
gravitino picks up a phase, $-1$.) So the last
orbifolding breaks supersymmetry. The $\field{Z}_3$ orbifold
\ref{trans}-\ref{TT} is supersymmetric and so has no tachyons in
the spectrum. Only the last orbifold by \ref{tQ}\ breaks
supersymmetry, but this transformation acts freely due to the
translation in the third $T^2.$ So the mass terms in the twisted
sectors associated with \ref{tQ}\ will not have tachyons at large radius.

\subsection{Moduli stabilization}

The projectors in this model are almost the same as those of DeWolfe et al.  They
differ only in their action on space-time fermions.  As a result, the
moduli are the same as those of \cite{dewolfe}.  Moreover, the fluxes are the
same as well.  The moduli include the diagonal components of the metric
\begin{equation}\label{gam}
g_{i\bar i}\equiv\gamma_i.
\end{equation}
The forms include
a set of 2-forms
\begin{equation}\label{tf}
\omega_i\propto dz^i\wedge dz^{\bar i}
\end{equation} invariant under \ref{TT}, \ref{tQ} and odd under the spacetime reflection \ref{sig}.
There are 3 NSNS 2-form moduli $b_i$ with $B_2=b_i\omega^i$ that
combine with \ref{gam}\ to form 3 complex moduli. The holomorphic
3-form invariant under \ref{TT}, \ref{tQ} is again \ref{omeg}\
with the $O6$ plane wrapping the even cycle $\alpha_0$. Finally,
the dilaton joins the RR 3-form axion $\xi\alpha_0$ to form a
fourth complex scalar. As a result, the potential for the
untwisted moduli is again given by eqn. \ref{pot}\ with minima
given by eqn. \ref{min}. Also the stability analysis for the
twisted moduli carries through in the exact same way as in
\cite{dewolfe}.

To recapitulate, we used a flux compactification
to stabilize all the moduli, so that by taking a large enough CY
orientifold we can ensure that there are no tachyons in the spectrum, and this non-supersymmetric compactification is perturbatively stable.
An alternative model based on a $Z_4$ orbifold is described in appendix A.

\subsection{Quantum corrections}

One might worry that quantum corrections in such
non-supersymmetric models would be large due to the lack of
supersymmetry, either destabilizing the would-be AdS vacua, or
simply making any analysis impossible. After all, there are no
longer non-renormalization theorems, and the effective cutoff for
loop diagrams is at least as large as the Kaluza-Klein scale.
But, as in the supersymmetric case, higher order corrections
appear to be suppressed by powers of the four form flux, as we now
show.

Classically, the potential for the volume moduli $v_i$ behaves as
\begin{equation}
V_0 = {e^{2 \phi} \over v^6 }
\end{equation}
at the minimum.  It is important to note that this result is expressed in
Planck units, i.e. it is multiplied by $M_p^4$.  If we translate this into the mass
of the scalar, we need to recall that the scalar kinetic term is proportional to
$1/v^2$, so the tree level mass behaves as
\begin{equation}
m_0^2 = {e^{2 \phi} \over v^6} M_p^2 = {1 \over v^3} M_s^2.
\end{equation}

Let's compare with what we might expect at one loop.
We can estimate these effects in two equivalent ways.  First, consider a Casimir computation.
In string frame, this will give a result of the form
\begin{equation}
\delta V \sim {1 \over R^4} \sim {1 \over v^2} M_s^4.
\end{equation}
so
\begin{equation}
\delta m^2 \sim e^{2 \phi}{1 \over v^5}{M_s^2}.
\end{equation}
The, the loop correction is suppressed by $1/v^3 \sim \vert e_i
\vert^{-3/2}$ and for large four-form flux, the corrections are
under control\footnote{Alternatively, we can consider a direct
computation of the mass.  For this we can work with canonically
normalized fields.  There is a one loop correction to the mass
involving emission of a graviton.  This correction is
quadratically divergent; we expect that it is cut off at $1/R$.
So the size of the mass correction is $\delta m^2 \approx G_N {1
\over R^4}=e^{2 \phi}{1 \over v^5}M_s^2$ as above.}.

This analysis is arguably naive.  It is not known
how to perform a string perturbation expansion for these configurations,
and the work of \cite{banks} suggests that this might not be possible.
We simply note here that the low energy theory, viewed as a cutoff
field theory, appears to be under control.

This example shows that there are models with badly broken
supersymmetry but where classically all the moduli are stabilized.
Our basic strategy was to note
that the Scherk-Schwarz projection does not
change the classical
equations for the bosonic fields,
and these can be shown in some cases to be
subject to small quantum corrections. In appendix \ref{ze4}
we present a closely related $\field{Z}_4\times \field{Z}_4$ CY
orientifold, where again we are able to break susy at the KK scale
while retaining perturbative stability.

\section{Conclusions:  Potential Applications}\label{concl}

We have seen that it is possible to generate a wide array of non-supersymmetric
string configurations using a generalized notion of Scherk-Schwarz projection.
Models of this type should be useful theoretical laboratories for studying
a number of questions.  To conclude, we mention some areas for
further study.

\subsection{Non-Perturbative Instabilities with Stabilized Moduli}

Kaluza-Klein spaces with standard Scherk-Schwarz boundary conditions suffer
from a non-perturbative instability to decay into a ``bubble of
nothing'', even in the regime where the size of the orbifold is
big and there are no winding tachyons. The instability was derived
by Witten in \cite{wittenkk}\ using an analytical continuation of
the Schwarzschild solution.  In most applications, the significance
of this solution is obscure for a variety of reasons, perhaps the most
important being that the classical, non-supersymmetric string or field
theory Kaluza-Klein solutions are already unstable
in perturbation theory.

The IIA models offer promise of studying this question, as all
of the moduli are stable.  Presumably, the Witten bounce in this
case is associated with the non-contractible loops introduced by the Scherk-Schwarz
projection.  Efforts to construct such solutions
in states with all moduli fixed will be reported elsewhere.

\subsection{The Landscape}

Non-supersymmetric models with fluxes of the type described here
provide an arena for examining landscape statistics.   There are
a variety of questions which one might try to address.  These include
raw counting, e.g. trying to compare numbers of supersymmetric
and non-supersymmetric states with various properties.  In the IIA
case, for large fluxes, the states, classically, are AdS, and small
quantum effects will not change this.  We have noted that non-supersymmetric
IIB theories are more challenging to study than their supersymmetric
counterparts.  The complex structure moduli are fixed, classically,
as in the supersymmetric case, but the Kahler moduli will have potentials
already in perturbation theory, and stationary points will typically
lie at small radius.  One would expect, however, that such states
would exist, and that some will be dS, some AdS.  More refined
questions include counting of states with discrete symmetries and
with exponentially large warping (we thank
Steve Giddings for a discussion which led us to consider
this application).  Studies of supersymmetric vacua\cite{douglasdenef},
for example, having indicated that exponential warping occurs in a substantial
fraction of states.  The models described here with most or all moduli
stabilized provide a further laboratory for investigating this question.
It seems likely that one will again find that exponential warping is common.
This might suggest that non-supersymmetric vacua with warping are more likely
solutions of the hierarchy problem than supersymmetric ones.  Of course,
whether this can be reconciled with other facts of low energy physics
is an important question.

\subsubsection*{\bf Acknowledgements:} We thank T. Banks, S. Giddings, O. DeWolfe, S. Kachru
and E. Silverstein for conversations.  We are particularly grateful
to Volker Braun for pointing out to us an error in the first version
of this manuscript, and correcting a misconception of ours.
This work supported in part by the U.S.
Department of Energy.  M.D. thanks the Kavli Institute for
Theoretical Physics for hospitality.

\section*{Appendix A -Generalized Scherk-Schwarz in a $\field{Z}_4^3$ example.}\label{ze4}

In this section we consider a $\field{Z}_4\times \field{Z}_4$
orbifold. We start from a $T^6$ defined by
\begin{equation}
z_j\sim z_j+1\sim z_j+i\quad j=1,2,3
\end{equation} and mod out by a $\field{Z}_4\times \field{Z}_4$ symmetry
\begin{equation}\label{AA}\begin{split}
z_i\rightarrow &A_i^{\ j} z_j {\rm\quad with\quad } A=\left (
\begin{matrix}i & 0 & 0 \cr 0  & -i & 0 \cr 0 & 0 & 1
\end{matrix}\right ) {\rm\quad and\quad}\\ z_i\rightarrow &B_i^{\
j} z_j {\rm\quad with\quad } B= \left ( \begin{matrix}i & 0 & 0
\cr 0  & 1 & 0 \cr 0 & 0 & -i\end{matrix}\right ).
\end{split}\end{equation}
There are many fixed points (some of which are only fixed under $Z_2$ subgroups).
This orbifold model is consistent with the same orientifold
projection employed in \cite{dewolfe}\ and described here in eqn.
\ref{ori}. The $\field{Z}_4\times \field{Z}_4$ torus orientifold
keeps precisely the same untwisted moduli as in \cite{dewolfe}.
It is fairly straightforward to check that the stabilization
analysis of the untwisted moduli for this model precisely follows
the analysis carried by DeWolfe et.al in \cite{dewolfe}.  As a
result, the potential for the untwisted moduli has a form
identical to that of the $\field{Z}_3$ case (reproduced here in
eqn. \ref{pot}), with similar minima given by eqn. \ref{min}.

To  achieve Scherk-Schwarz compactification, one can now
orbifold this model by a freely
acting $\field{Z}_4$
\begin{equation}\label{z4ssf}
\begin{split}
z_i\rightarrow C_i^{\ j} z_j + a_i{\rm\quad with\quad } C=\left (
\begin{matrix}e^{\pi i\over 2} & 0 & 0 \cr 0  & e^{3\pi i\over 2}
& 0 \cr 0 & 0 & 1 \end{matrix}\right ) {\rm\quad and\quad}\ a=
\frac{1+i}{2}\left ( \begin{matrix}1 \cr 1\cr1\end{matrix}\right
).
\end{split}
\end{equation}
For this $\field{Z}_4^3$ CY orientifold, just like in the
$\field{Z}_3^2$ case, susy is broken at the KK scale, and
the fluxes can stabilize the geometry to a safe size where the
spectrum is free of tachyons\footnote{Again, modulo possible surprised from blow up modes which where not analyzed.}. Again, an effective
field theory analysis suggests that the quantum corrections are
small.

To complete the stabilization analysis one needs to study the stabilization of blow up modes at the various fixed points of this orbifold.
The analysis is more complicated in this case, because there are many more sectors to consider, with a more intricate
pattern of couplings.  We see no reason to believe that one cannot fix the moduli in a controlled fashion, as
in the $Z_3^2$ case \cite{dewolfe}.  A complete analysis will be presented elsewhere\cite{ms}.

\section*{Appendix B - Type $0$ and Scherk-Schwarz in RNS.}\label{apb}
\subsection*{$\ 1.\qquad$Projections in operator language}

The usual treatment of type $0$ string theory \cite{sw}, is based
on modular invariance of the partition function. One of the
interesting features of type $0$ models is their duality relation
with Scherk-Schwarz compactifications \cite{rohm}, or
compactification on ``twisted circles'' as they are sometime
called, when the radius of the twisted circle goes to zero. This
duality relies on the existence of winding tachyons in
Scherk-Schwarz compactifications due to a reversal of the GSO
projection in odd winding sectors, as explained e.g. in
\cite{atickwitten}. In this appendix we present an equivalent
treatment in operator language. The reason for doing so is that in
this language the modified GSO projection in odd winding sectors
as well as the relation to type $0$ string theory is particularly
transparent. The basic premise we use is that to ensure the
consistency of a worldsheet sigma model as a string background one
must make sure that
\begin{itemize}
\item A) All the operators of the worldsheet sigma model are mutually local.
\item B) The operator algebra is closed.
\item C) Left-right level matching in imposed.
\end{itemize}
Projections are carried out by demanding that an operator
$\mathcal{O}$ that enforces the projection via its OPE with the
rest of the operator algebra, is part of the worldsheet SCFT.
Condition A projects on invariant states while condition B will
then generate twisted sectors.

For example, the type II GSO projection is arrived at by demanding mutual locality and closure with the spacetime supercharges.
Similarly, if we want to describe a compactification of type II on a circle \ref{arr}\ we add to the SCFT the two operators
\begin{equation}\label{wind}
{R}(z,\bar z)\equiv e^{i\frac{R}{\alpha^{\prime}}\bigl(X(z)-\tilde
X(\bar z)\bigr)}
\end{equation} and
\begin{equation}\label{mom}
 M(z,\bar z)=e^{i\frac{1}{R}\bigl(X(z)+\tilde X(\bar
z)\bigr)}
\end{equation} and follow the steps mentioned above. This will project on the correct momentum lattice, via OPE with $e^{ip_LX(z)}e^{ip_R\tilde X(\bar z)}$, and create the winding and momentum sectors.

\subsection*{$\ 2.\qquad$The type $0$ projection}

Let us examine what happens when we start from a type II superestring on $\field{R}^{1,9}$.
The holomorphic spacetime supercharges are built out of the bosonized fermions $H_1,\dots,H_5$ and the reparameterization superghost $\varphi$ as
\begin{equation}\label{sss}
Q_{\alpha}=\oint e^{-{\varphi\over 2}}e^{{i\over 2}(\epsilon_1H_1+\dots+\epsilon_5H_5)}
\end{equation} where $\alpha=[\epsilon_1,\dots,\epsilon_5]$ is a spinor index of $SO(1,9)$ and each $H$ bosonizing a pair of fermions, say $\psi^{1,2}$, according to the familiar formulas
\begin{equation}
\psi^1=\sqrt{2}\cos H,\quad \psi^2=\sqrt{2}\sin H,\quad \psi^{\pm}\equiv\frac{1}{\sqrt{2}}\bigl(\psi^1\pm i\psi^2\bigr)=e^{\pm iH}
\end{equation}

Demand now that the worldsheet CFT includes the following operator\footnote{Note that this is not a physical operator and should not be confused with a very similar physical operator in the $(-1,-1)$ picture
given by $e^{-\varphi-\tilde\varphi}\psi^+\tilde\psi^-$.}$^,$\footnote{In fact, the closure of the OPE algebra shows that it does not really matter if we use any combination of the form $e^{{i}\bigl(H_k(z)\pm\tilde H_l(\bar z)\bigr)}$ with $k\neq l$. We thus refer to this operator simply as \ref{tee}\ without specifying the labels $k,l$ or the relative sign.}
\begin{equation}\label{tee}
T(z,\bar z)=e^{{i}\bigl(H(z)-\tilde H(\bar z)\bigr)}.
\end{equation}

The fact that $T(z,\bar z)$ is not a chiral operator suggests that the resulting theory will not respect the chiral GSO projection. In fact, we will now show that the resulting theory is the well known type $0$ model.
Using the standard relation
\begin{equation}\label{ope}
:e^{iaX(z)}::e^{ibX(w)}:=(z-w)^{ab}:e^{i\bigl(aX(z)+bX(w)\bigr)}:
\end{equation} it is clear that all the operators in the $(NS,NS)$ sector are mutually local with $T(z,\bar z)$.
However, spin fields $S_{\alpha}\equiv e^{{i\over
2}(\epsilon_1H_1+\dots+\epsilon_5H_5)}$ (as well as their
anti-holomorphic counterparts) have a square root branch cut in
their OPE with $T(z,\bar z)$. This means that in the resulting
theory there are no operators in the $(R,NS),\ (NS, R)$ sectors,
and hence no spacetime fermions. Moreover, since the spacetime
supercharges \ref{sss}\ have been projected out, we no longer have
the usual chiral GSO projection and in particular the $(NS,NS)$
vacuum corresponding to a tachyon in spacetime survives the
projection.

What about the $(R,R)$ sector? Starting from a type II superstring model we see that any pre-existing $(R,R)$ vertex operator survives the projection
\begin{equation}\label{tss}
T^{(k)}(z,\bar z)\cdot S_{\alpha}(w)\tilde S_{\dot\alpha}(\bar w)\sim (z-w)^{\epsilon_1/2}(\bar{z}-\bar w)^{-\tilde\epsilon_1/2}\dots=|z-w|^{\epsilon_1}(\bar z-\bar w)^{-(\tilde\epsilon_1+\epsilon_1)/2}\dots
\end{equation} and since $(\tilde\epsilon_1+\epsilon_1)/2\in\field{Z}$ there is no branch cut.
However, The $(R,R)$ spectrum is doubled. The worldsheet fermions act on the spinfields as spacetime $\Gamma$ matrices and so the ``twisted sectors'' which one gets on the RHS of \ref{tss}\ are $(R,R)$ vertex operators where the spacetime chiralities are flipped {\it both} for the left and right movers.

To summarize\footnote{Following the notations used by Polchinski in chapter 10 pages 26-27 of \cite{pol}}, If we started from type IIA/B with the sectors
\begin{itemize}
\item IIA: $(NS+,NS+), (R+,NS+), (NS+,R-), (R+,R-)$
\item IIB: $(NS+,NS+), (R+,NS+), (NS+,R+), (R+,R+)$,
\end{itemize} demanding that \ref{tee}\ is part of the worldsheet CFT projects the theory onto
\begin{itemize}\label{zerosect}
\item 0A: $(NS+,NS+), (NS-,NS-), (R-,R+), (R+,R-)$
\item 0B: $(NS+,NS+), (NS-,NS-), (R-,R-), (R+,R+)$
\end{itemize} which is exactly the type $0$ projection.

\subsection*{$\ 3.\qquad$Scherk-Schwarz projection}

A worldsheet operator realization of Scherk-Schwartz boundary
conditions for type II superstrings can be achieved by adding to
the type II worldsheet algebra the {\it combined} operator
\begin{equation}\label{ppp}
P(z,\bar z)\equiv \bigl[R\cdot T\bigr](z,\bar z).
\end{equation} where $R, T$ where defined in \ref{wind}, \ref{tee}.
In words, this operator insists that when we create a winding mode
we must accompany each winding with the operator that performs the
type $0$ projection. Let us examine the effect of inserting
\ref{ppp}. Mutual locality with \ref{ppp}\ projects out the zero
momentum vertex operators \ref{sss}\ thus breaking spacetime
supersymmetry.

However, this model does have spacetime fermions. Working out the OPE of \ref{ppp}\ with an $(R,NS)$ vertex operator that carries some momentum along the circle ($p_{L,R}$ where defined in \ref{plpr})
\begin{equation}\label{rnsvop}
\mathcal{V}(z,\bar z)\sim e^{\frac{i}{2}(\epsilon_1H_1+\dots+\epsilon_5H_5)}e^{ip_LX}e^{ip_R\tilde X}
\end{equation} we get the following (possible) branch cut in $z$
\begin{equation}
P(z,\bar z)\cdot \mathcal{V}(z,\bar z)\sim z^{m+\frac{1}{2}}\cdots
\end{equation} with the conclusion that in order for this vertex operator to be projected in (giving a physical spacetime fermion state) we must restrict
\begin{equation}
m\in\field{Z}+\frac{1}{2}.
\end{equation}
Physically, this is just the familiar Scherk-Schwarz projection
where fermions have anti-periodic boundary conditions (and
therefore half integral momenta) along the circle.

So far we have checked mutual locality with \ref{ppp}\ which in
the language of orbifolds, is the projection on invariant states.
The twisted sectors arise by closing the OPE algebra with
\ref{ppp}. Let us examine the bosonic spectrum (suppressing the
$(R,NS), (NS,R)$ sectors). At the $w=0$ sector we have the usual
type II spectrum $(NS+,NS+), (R+,R\pm)$ depending on whether we
are in type IIA/B. For example we have the graviton vertex
\begin{equation}\label{grav}
G^{ij}(z,\bar z)\equiv e^{-\varphi-\tilde\varphi}\psi^i\tilde\psi^j.
\end{equation}
Going to the first twisted sector $w=1$ we have to perform the OPE not with $R$ (eq. \ref{wind}) as if we are doing a ``normal'' circle but with $P$ (eq. \ref{ppp}). The leading singular term will have the $T$ in $P$ contract against the fermions in \ref{grav}\ leaving us with the vertex operator
of a {\it winding tachyon}
\begin{equation}
e^{-\varphi-\tilde\varphi}\cdot \widehat{R}
\end{equation}

In the $(R,R)$ sector, the effect of $T$ is to flip the
chiralities on {\it both} the left and right movers, so we end up
in the $w=1$ sector\footnote{This phenomenon is sometime referred
to by saying that the GSO projection is reversed in the odd
winding sectors, but one should be careful about this phrasing
because in type $0$ theories the GSO is not chiral and the
reversing happens in each chirality separately.} with $(NS-,NS-),
(R-,R\mp)$. It is straightforward to see that this picture
persists to higher winding sectors:
\begin{itemize}
\item $w=2w^{\prime}\ $: $(NS+,NS+), (R+,R\pm)$
\item $w=2w^{\prime}+1\ $: $(NS-,NS-), (R-,R\mp)$.
\end{itemize} This demonstrates the reversal of the GSO projection in odd winding sectors.

\subsection*{$\ 4.\qquad$Scherk-Schwarz on a vanishing circle.}
Let us see what happens in the limit that $R\rightarrow 0$.
Naively in this limit the operator \ref{wind}\ is just the
identity so $P$ becomes  the type 0 projection $T$ and we expect
to get a type 0 model. This sloppy argument actually gives the
correct answer. In this limit all the spacetime fermions become
very massive because they can not have zero modes along the
circle. Furthermore, given an integer $w^{\prime}$, the successive
winding sectors $w=2w^{\prime},2w^{\prime}+1$ are almost
degenerate when $R<< \alpha^{\prime}=2$
\begin{itemize}
\item $w=2w^{\prime}\ $: $(NS+,NS+), (R+,R\pm)\quad\longrightarrow m^2=\bigl(\frac{2w^{\prime}R}{2}\bigr)^2-1$
\item $w=2w^{\prime}+1\ $: $(NS-,NS-), (R-,R\mp)\quad\longrightarrow m^2=\bigl(\frac{(2w^{\prime}+1)R}{2}\bigr)^2-1$.
\end{itemize}
Neglecting the difference $\Delta m^2=R^2/4$ we see that the
spectrum in those two successive sectors combined is that of {\it
the $w^{\prime}$ winding sector of the corresponding type $0$
theory compactified on a circle with twice the radius}
$R^{\prime}=2R$. In the limit $R\rightarrow0$ this become exact
and it makes more sense instead of talking about the type 0A/B
model on a vanishing circle to describe the model as type 0B/A in
uncompactified spacetime.

\end{document}